\def\year{2021}\relax
\documentclass[letterpaper]{article} 
\usepackage{aaai21}  
\usepackage{times}  
\usepackage{helvet} 
\usepackage{courier}  
\usepackage[hyphens]{url}  
\usepackage{graphicx} 
\usepackage{natbib}  
\usepackage{caption} 
\usepackage{booktabs}
\usepackage{amsfonts}
\usepackage{amsthm}
\theoremstyle{definition}
\newtheorem{definition}{Definition}
\usepackage{listings}
\usepackage{xcolor}
\usepackage{amsmath}
\usepackage{enumitem}
\usepackage{xspace}
\usepackage{subcaption}

\newcommand{\hide}[1]{}

\newcommand{\ours}{CCAG\xspace} 
\newcommand{\oursg}{CCAG$_g$\xspace} 
\newcommand{\oursn}{CCAG$_n$\xspace} 
\newcommand{\oursb}{CCAG$_{b}$\xspace} 
\newcommand{\oursp}{CCAG$_{p}$\xspace} 
\newcommand{\oursr}{CCAG$_{r}$\xspace} 
\newcommand{\oursgs}{CCAG$_{gs}$\xspace} 
\newcommand{\oursng}{CCAG$_{ng}$\xspace} 
\newcommand{\ourspe}{CCAG$_{pe}$\xspace} 

\graphicspath{{fig/}}

\title{Code Completion by Modeling Flattened Abstract Syntax Trees as Graphs}
\author {
    Yanlin Wang,\textsuperscript{\rm 1}
    Hui Li\textsuperscript{\rm 2}\thanks{Hui Li is the corresponding author.}\\
}
\affiliations {
    \textsuperscript{\rm 1} Microsoft Research Asia \\
    \textsuperscript{\rm 2} School of Informatics, Xiamen University \\
    yanlwang@microsoft.com, hui@xmu.edu.cn
}

\begin{document}
\maketitle

  
\begin{abstract}
Code completion has become an essential component of integrated development environments. Contemporary code completion methods rely on the abstract syntax tree (AST) to generate syntactically correct code. However, they cannot fully capture the sequential and repetitive patterns of writing code and the structural information of the AST. To alleviate these problems, we propose a new code completion approach named CCAG, which models the flattened sequence of a partial AST as an AST graph. CCAG uses our proposed AST Graph Attention Block to capture different dependencies in the AST graph for representation learning in code completion. The sub-tasks of code completion are optimized via multi-task learning in CCAG, and the task balance is automatically achieved using uncertainty without the need to tune task weights. The experimental results show that CCAG has superior performance than state-of-the-art approaches and it is able to provide intelligent code completion.
\end{abstract}


\section{Introduction}
\label{sec:intro}

Code completion, which provides code suggestions for developers, is one of the
most attractive features in integrated development environments (IDEs).
According to the study of~\citet{MurphyKF06}, users of Eclipse IDE
used the code suggestion of Eclipse as much as the common editing commands
(e.g., copy and paste) since it reduces the required amount of typing
and eliminates typos. 

\citet{HindleBSGD12} firstly reduce code
completion to a natural language processing (NLP) problem. 
Thereafter, many
researchers leverage NLP techniques to design code completion
engines~\citep{AllamanisBDS18,LeCB20}. Early works generate code as a sequence
of code tokens~\citep{HindleBSGD12}. However, directly modeling tokens of code
sequences sometimes fails to produce syntactically correct
code~\citep{BrockschmidtAGP19}. Recent works~\citep{LiuWSGS16,LiWLK18}
alleviate this issue by using the target language's grammar to generate
abstract syntax trees (ASTs) which are syntactically correct by construction.
They show that AST based code completion, which contains value prediction and
type prediction as sub-tasks, can provide more intelligent code suggestions,
and it has been adopted in several IDEs (e.g., Visual Studio Code
IDE~\citep{SvyatkovskiyZFS19}). In this paper, the term ``code completion''
refers to AST based code completion.

\begin{figure*}[!t]
\begin{center}
\includegraphics[width=1\textwidth]{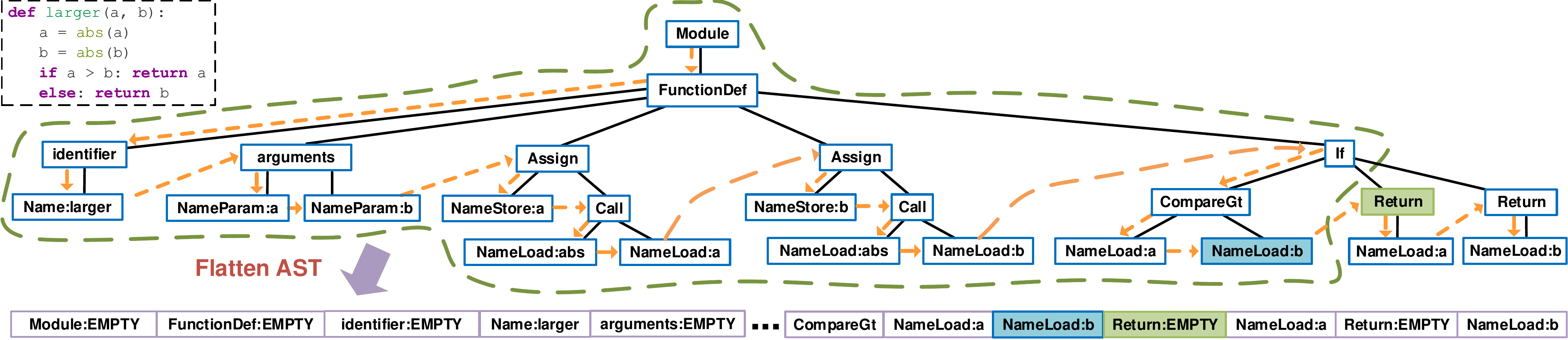}
\caption{The AST of a Python function. 
The orange dashed arrows show the traversal for the flattening. The green dashed line indicates a partial AST with \texttt{NameLoad:b} being the \emph{right-most} node and \texttt{Return} (i.e., \texttt{Return:EMPTY} in the flattened AST) being the next node to predict. 
}
\label{fig:ast_example}
\end{center}
\end{figure*}

To model the tree structure of AST, most code completion methods opt to flatten the AST using
pre-order depth-first traversal~\citep{LiuWSGS16,LiWLK18}. Then, powerful deep
learning techniques (e.g., LSTMs and Transformer)  can be adopted for learning
the representation of the flattened AST sequence for later use in code
completion. In addition to code completion, recent works on learning program representations~\citep{AllamanisBK18,BrockschmidtAGP19} for other downstream tasks (e.g., variable prediction and hole completion) have shed some light on the
benefits of modeling an AST as a graph of which the representation can be
learned using Graph Neural Network (GNN)~\citep{WuPCLZY20}. 

However, neither of the two paradigms can fully model ASTs for the code
completion task since they neglect the  sequential and repetitive patterns of
humans on writing code, and the structural information of AST: (1) Firstly, when
writing code, the skeleton (i.e., function declaration) is often written first
and other statements in the body of the function are written one by one just
like the pre-order depth-first traversal of AST~\citep{abs-2003-08080}.
Such sequential information is important to code completion. (2) Moreover,
code completions are surprisingly repetitive~\citep{HellendoornPGB19}. For
example, the study of \citet{abs-2004-05249} on a codebase from GitHub shows
that there is a 59.8\% probability that any keyword, identifier, or literal
repeats one of the previous 100 tokens. Capturing such a repetitive pattern can
enhance code completion. (3) Lastly, the structural information of AST
provides strong indications on the dependencies between linked nodes which
should be considered. Simple sequential modeling ignores
the repetitive pattern and the structural information while the vanilla graph
based modeling neglects the sequential and repetitive patterns. In
addition to the problem of AST modeling, some code
completion methods~\citep{LiuWSGS16,LiuLWXFJ20} model the sub-tasks (i.e., value prediction and type
prediction) via multi-task learning~\citep{abs-2004-13379}, but the task
weights are manually set. Therefore, these
methods suffer from the task imbalance which will impede proper training in
multi-task learning~\citep{ChenBLR18,abs-2004-13379}.

To address the problems mentioned above, we propose an effective \underline{C}ode \underline{C}ompletion method by modeling flattened \underline{A}STs as \underline{G}raphs (\ours for short). Compared to previous methods, the contributions of our work are:
\begin{itemize}[topsep=2pt,itemsep=0pt,leftmargin=10pt]
\item \ours models a flattened AST derived from the pre-order depth-first traversal of a partial AST as a graph, and it is tailored to include the sequential and repetitive patterns of writing code and the structural information of the AST. 
\item \ours uses our proposed AST Graph Attention Block (ASTGab) comprised of three different attention based layers. Each layer captures differing dependencies among AST nodes. ASTGab is further enhanced with the residual connection so that multiple ASTGabs can be stacked to improve the performance.
\item \ours adopts an uncertainty based method to automatically balance the two sub-tasks of code completion in multi-task learning without the need to tune task weights.
\end{itemize}
We conduct extensive experiments on benchmark data for evaluating code completion. Results show that \ours surpasses state-of-the-art code completion approaches.


\section{Preliminary}
\label{sec:pre}

Any programming language has an explicit context-free grammar (CFG), and it can
be used to parse source code into an AST which represents the abstract syntactic
structure of source code. An AST is a tree where each non-leaf
node corresponds to a \emph{non-terminal} in the CFG specifying structural
information (e.g., \texttt{ForStatement} and \texttt{IfStatement}). Each leaf node corresponds to a \emph{terminal} (e.g.,
variable names and operators) in the CFG
encoding program text. An AST can be converted back into source code easily. Fig.~\ref{fig:ast_example} provides an example of the AST for a python function. We can see that each non-leaf node contains a type attribute (e.g., \texttt{Module}) and each leaf node contains a type attribute and a value attribute (e.g., \texttt{NameLoad:a} means that the type is \texttt{NameLoad} and the value is \texttt{a}). 

Code completion consumes a partial AST as the input:
\begin{definition}[Partial AST~\citep{LiuWSGS16}]
Given a complete AST $T$, a partial AST is a subtree $T'$ of $T$, such that for each node $n$ in $T'$, its left sequence $L_T(n)$ with respect to $T$ is a subset of $T'$, i.e., $L_T(n)\subseteq T'$. Here, the left sequence $L_T(n)$ is defined as all the nodes that are visited earlier than $n$ in the pre-order depth-first traversal sequence of $T$.
\end{definition}
For example, the green dashed line in Fig.~\ref{fig:ast_example} illustrates a partial AST. Following \citet{LiWLK18}, we append \texttt{EMPTY} as the value to each non-leaf node when flattening the AST. 
The formal definition of code completion is as follows:
\begin{definition}[Code Completion~\citep{LiuWSGS16}]
Each partial AST $T'$ has one \emph{right-most} node $n_R$, and all other nodes of $T'$ form its left sequence $L_T(n_R)$. We call the next node after $n_R$ in the pre-order depth-first traversal sequence of the complete AST $T$ as the next node following $T'$. Given a partial AST $T'$, the task of code completion is to predict the value and the type of the next node following $T'$.
\end{definition}
For the partial AST shown in Fig.~\ref{fig:ast_example}, the right-most node is \texttt{NameLoad:b} and the next node to predict is \texttt{Return} (i.e., \texttt{Return:EMPTY}). A successful model should give both the value \texttt{EMPTY} and the type \texttt{Return} as predictions.

The traversal order of ASTs in code completion\footnote{Note that some works use in-order depth-first traversal to define the problem~\citep{LiuWSGS16,LiWLK18,LiuLWXFJ20}. But the AST examples in their papers and the code completion task in their experiments use pre-order depth-first traversal.} is consistent with the way that developers implement a function: the function declaration is often written first and then other statements are written one by one~\citep{abs-2003-08080}.


\section{Learning to Complete Code with \ours}

\begin{figure}[!t]
  \centering
  \includegraphics[width=0.98\linewidth]{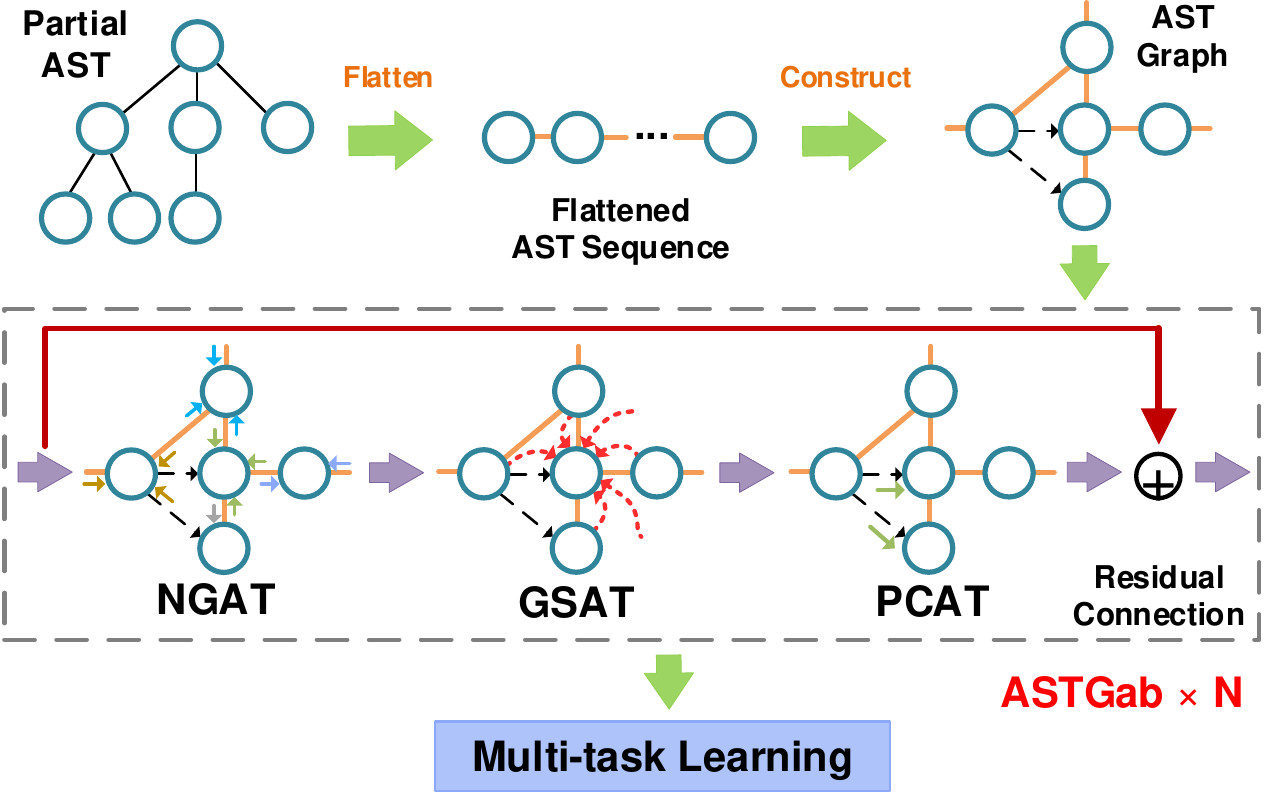}
  \caption{Overview of \ours. In graphs, orange lines show the traversal for flattening the AST or the node-node edges, and black dashed lines are parent-child edges.} 
  \label{fig:model}
  \vspace{-10pt}
\end{figure}

In this section, we will describe the details of \ours. Fig.~\ref{fig:model} provides an overview of \ours.

\subsection{Program Representation}
\label{sec:pro_rep}

\begin{figure}[!t]
  \centering
  \includegraphics[width=0.92\linewidth]{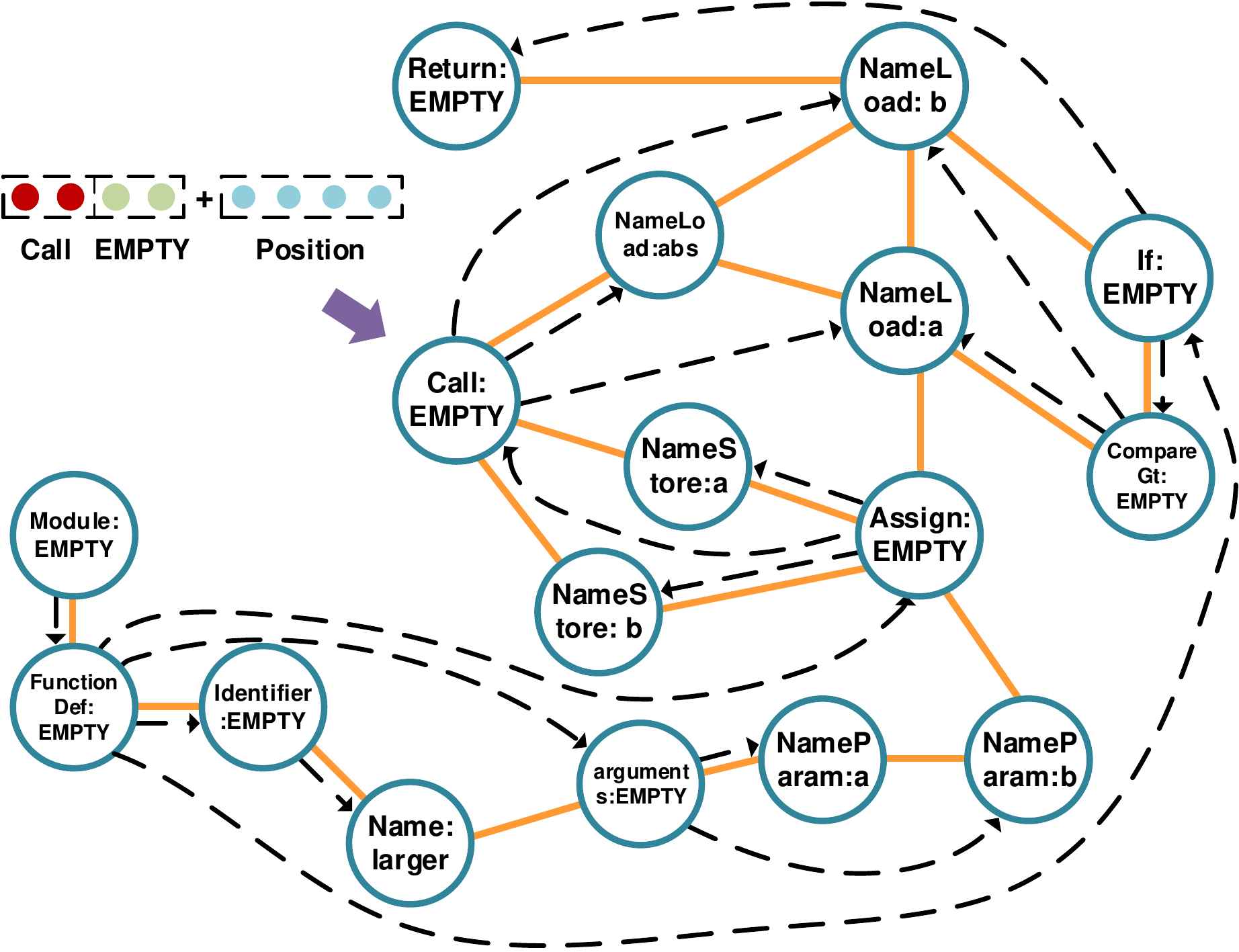}
  \caption{The AST graph in \ours for the partial AST in Fig.~\ref{fig:ast_example}. Orange lines show the node-node edges and black dashed lines indicate the parent-child edges.} 
  \label{fig:ast_graph}
  \vspace{-10pt}
\end{figure}

\ours models
each flattened AST sequence of a partial AST as an AST graph. Fig.~\ref{fig:ast_graph} shows how
\ours represents the partial AST in Fig.~\ref{fig:ast_example} as a graph. Duplicated nodes in the flattened AST are merged into one 
node in the graph. Each \emph{node-node} edge in the graph indicates
that the two linked nodes are adjacent in the flattened AST sequence. Node-node edges are undirected, allowing information propagation in both directions. The weight of a node-node edge is the frequency of the
occurrence of the edge in the corresponding flattened AST sequence.

However, flattening a partial AST into a sequence may result in the information
loss of the tree structure. Following \citet{LiWLK18}, we record
the parent node of each node in original ASTs since parent-child information can
help the model learn the hierarchical structure of the AST. Then, we add
\emph{parent-child} edges to the AST graph. Parent-child edges are directed (from
parent to child) and unweighted to retain the structure. Each node may have more than one parent
node. 

One remaining issue is that the positional information of each node in the
flattened AST sequence is missing in the graph, since repeated AST nodes are
merged into one node in the graph. To remedy it, we record
the distance between the last occurrence of each node and the right-most node in the flattened AST sequence as the positional encoding. For instance, assuming that the flattened AST sequence is $\{n_1, n_2, n_3, n_2\}$ with $n_2$ being the right-most node, the position embeddings $\mathbf{p}_1, \mathbf{p}_2, \mathbf{p}_3$ for $n_1$, $n_2$ and $n_3$ are vectors with all dimensions being 3, 0 and 1, respectively. Note that the position embeddings are fixed and will not be updated.

For a node $n_i$, we embed its value and type into two
separate spaces. $\mathbf{v}_i, \mathbf{t}_i\in\mathbb{R}^d$, $\mathbf{p}_i\in\mathbb{R}^{2d}$ are its value embedding vector, type embedding
vector and position embedding vector, respectively. 
The representation $\mathbf{h}_i$ of $n_i$ is as follows: 
\begin{equation}
\small
\label{eq:emb}
\mathbf{h}_i = ReLU\big(\mathbf{W}^{(p)}(\big[\mathbf{t}_i \left|\right| \mathbf{v}_i\big] + \mathbf{p}_i)+\mathbf{b}^{(p)}\big),
\end{equation}
where $\left|\right|$ indicates the concatenation operation, $\mathbf{W}^{(p)}\in\mathbb{R}^{d\times 2d}$ is a parameter matrix and $\mathbf{b}^{(p)}$ is the bias vector.

Since we first flatten the partial AST and then derive the AST graph from the flattened AST sequence, one may ask why using such indirect modeling. Here, we provide discussions on alternatives to justify the rationality of our design. In Sec.~\ref{sec:er}, we will also show that our design leads to superior performance than the alternatives.

\begin{itemize}[topsep=2pt,itemsep=0pt,leftmargin=10pt]
\item\textbf{Alternative 1: Directly modeling the original partial AST as a graph/tree}. In such a design, sequential information is missing, and repeated nodes may be located in different parts of the graph which makes it hard for the model to capture the repetitive pattern. A tree can be viewed as a special graph and has same issues as mentioned above.
\item\textbf{Alternative 2: Directly modeling the flattened AST sequence.} Directly modeling the flattened AST sequence also makes capturing the repetitive pattern harder since repeated nodes can be physically distant from each other in the sequence. Additionally, simple sequential modeling will result in the loss of the structural information of the partial AST. 
\end{itemize}

Differently, in \ours, the sequential pattern is incorporated by modeling the flattened AST sequence, the repetitive pattern can be captured via the merged repeated nodes and the weights of node-node edges, and the structural information is retained via the parent-child edges.

\subsection{AST Graph Attention Block (ASTGab)}

We adopt the idea of GNN to design an AST Graph Attention Block (ASTGab) for learning the representation of the AST graph. GNN is well-suited for learning AST graphs since it is designed to automatically extract features from the rich node connections in the graph data~\citep{WuPCLZY20}. 
The input to the ASTGab is the initial embeddings of all AST nodes in one AST graph derived from a partial AST. 
An ASTGab consists of three different layers
which extract node features at different levels, as shown in Fig.~\ref{fig:model}. The output is handled by a residual connection to overcome the difficulty of training deep neural networks.

\subsubsection{Neighbor Graph Attention Layer (NGAT).} The initial embeddings are first fed into a NGAT which extracts features from the \emph{first-order} neighborhood along the node-node edges in the AST graph. Strong dependencies typically exist between first-order neighbors. For instance, the comparison operators are normally followed by loading the variable. In the AST graph, this indicates a node-node edge between two AST nodes (e.g., the node-node edge between \texttt{CompareGt:EMPTY} and \texttt{NameLoad:a} in Fig.~\ref{fig:ast_graph}). 

Inspired by \citet{VelickovicCCRLB18}, we perform self-attention on every pairs of the first-order neighbors connected by the node-node edges in the AST graph and compute the first-order neighbor attention coefficient in NGAT:
\begin{equation}
\small
\label{eq:neighbor_att}
e_{i,j}^{(n)}=a(\mathbf{W}^{(n)}\mathbf{h}_i, \mathbf{W}^{(n)}\mathbf{h}_j,w_{i,j}),
\end{equation}
where $e_{i,j}^{(n)}$ shows the attention coefficient between first-order neighbors $i$ and $j$, $a$ is an attention mechanism,
$\mathbf{W}^{(n)}\in\mathbb{R}^{d\times d}$ is the shared weight matrix for all first-order node pairs, and $w_{i,j}$ is the weight of the node-node edge between $i$ and $j$. The first-order neighbor attention coefficients are then normalized across all the first-order neighbors of an AST node using softmax.
There are many choices for the design of $a$. We adopt a single-layer feedforward neural network followed by a LeakyReLU non-linearity unit with a negative input slope of 0.2 as \citet{VelickovicCCRLB18}. This is equivalent to the following expression:
\begin{equation}
\scriptsize
\label{eq:gat_a}
\alpha_{i,j}^{(n)}=\frac{\exp\bigg(LeakyReLU\Big(\mathbf{a}^{(n)^{T}}\big[\mathbf{W}^{(n)}\mathbf{h}_i \left| \right| \mathbf{W}^{(n)}\mathbf{h}_j \left| \right| w_{i,j} \big]\Big) \bigg)}{\sum_{k\in \mathcal{N}_i} \exp\bigg(LeakyReLU\Big(\mathbf{a}^{(n)^{T}}\big[\mathbf{W}^{(n)}\mathbf{h}_i \left| \right| \mathbf{W}^{(n)}\mathbf{h}_k \left| \right| w_{i,k} \big]\Big) \bigg)},
\end{equation}
where $\mathcal{N}_i$ indicates the set of first-order neighbors of node $i$ along node-node edges, and $\mathbf{a}^{(n)}\in\mathbb{R}^{2d+1}$ and $\mathbf{W}^{(n)}\in\mathbb{R}^{d\times d}$ are parameter vector and matrix, respectively. Then, the node feature for each AST node is updated as:
\begin{equation}
\small
\label{eq:gat_h}
\mathbf{h}_i^{(n)} = ReLU\big(\sum_{j\in \mathcal{N}_i}\alpha_{i,j}^{(n)}\mathbf{W}^{(a)}\mathbf{h}_j\big),
\end{equation}
where $\mathbf{W}^{(a)}\in\mathbb{R}^{d\times d}$ is the weight matrix.

As suggested by \citet{VelickovicCCRLB18}, we employ multi-head attention~\citep{VaswaniSPUJGKP17} in order to stabilize the learning of NGAT. Specifically, we perform $M$ independent transformations  in Eq.~\ref{eq:gat_h} and use the average of the $M$ results as the representation of each AST node:
\begin{equation}
\small
\mathbf{h}_i^{(n)} = ReLU\big(\frac{1}{M}\sum_{m=1}^{M}\sum_{j\in \mathcal{N}_i}\alpha_{i,j,m}^{(n)}\mathbf{W}^{(a)}_m\mathbf{h}\big),
\end{equation}
where $\mathbf{W}^{(a)}_m\in\mathbb{R}^{d\times d}$ is the parameter matrix for $m$-th head, and $\alpha_{i,j,m}^{(n)}$ is the normalized attention coefficient obtained by the $m$-th attention mechanism in Eq.~\ref{eq:gat_h}.

\subsubsection{Global Self-attention Layer (GSAT).} The second component GSAT captures the importance of each node to other nodes in the graph. Compared to NGAT, which focuses on the local neighborhood,  GSAT brings a global vision of the AST graph to \ours. Since the AST graph only contains unique AST nodes of which the number is much smaller than the original AST, we can feed the entire AST graph into GSAT, and then the self-attention mechanism is adopted to draw the global dependencies:
\begin{equation}
\small
\label{eq:gsat}
\mathbf{H}^{(g)} = softmax\big(\frac{(\mathbf{W}^{(k)}\mathbf{H}^{(n)})^{T}(\mathbf{W}^{(q)}\mathbf{H}^{(n)})}{\sqrt{d}}\big)\big(\mathbf{W}^{(v)}\mathbf{H}^{(n)}\big)^T,
\end{equation}
where $\mathbf{H}^{(n)}$ are the representations of AST nodes from NGAT, and $\mathbf{W}^{(k)}, \mathbf{W}^{(q)}, \mathbf{W}^{(v)}\in\mathbb{R}^{d\times d}$ are parameters.

\subsubsection{Parent-child Attention Layer (PCAT).} 
The third layer PCAT captures the structural information from parent-child edges and refine the node features from GSAT. Precisely, we adopt a two-layer attention mechanism in PCAT:
\begin{equation}
\small
\begin{aligned}
\mathbf{p}_i&=ReLU\big(\mathbf{W}^{(p1)}_1\mathbf{h}_i^{(g)}+\frac{1}{j\in\left|\mathcal{P}_i\right|}\mathbf{W}^{(p1)}_2\sum_{j\in\mathcal{P}_i}\mathbf{h}_j^{(g)}+\mathbf{b}^{(p1)}\big)\\
\mathbf{h}_i^{(p)}&=ReLU\big(\mathbf{W}^{(p2)}\mathbf{p}_i+\mathbf{b}^{(p2)}\big)
\end{aligned}
\end{equation}
where $\mathbf{h}_i^{(p)}$ is the representation for node $i$ outputted by PCAT and it incorporates the feature(s) from its parent node(s). $\mathcal{P}_i$ indicates the set of parent nodes of node $i$, $\left|\mathcal{P}_i\right|$ is the set size of $\mathcal{P}_i$, $\mathbf{h}_i^{(g)}\in \mathbf{H}^{(g)}$ is the representation for node $i$ from GSAT, $\mathbf{W}^{(p1)}_1, \mathbf{W}^{(p1)}_2, \mathbf{W}^{(p2)} \in \mathbb{R}^{d\times d}$ are learnable parameter matrices, and $\mathbf{b}^{(p1)}, \mathbf{b}^{(p2)}$ are bias vectors.

\subsubsection{Residual Connection.} As shown in Fig.~\ref{fig:model}, we can stack multiple ASTGabs to increase the non-linearities of \ours and the input to each ASTGab is the output from the previous ASTGab. However, deeper neural networks face the difficulty of training. Thus, we add a residual connection~\citep{HeZRS16} to each ASTGab to ease the difficulty: 
\begin{equation}
\begin{aligned}
\mathbf{r}_i&=ReLU\big(\mathbf{W}^{(r1)}\mathbf{h}_i^{(p)}+\mathbf{b}^{(r1)}\big)\\
\mathbf{h}_i^{(r)}&=\mathbf{W}^{(r2)}\mathbf{r}_i + \mathbf{b}^{(r2)} + \mathbf{h}_i
\end{aligned}
\end{equation}
where $\mathbf{W}^{(r1)}, \mathbf{W}^{(r2)} \in \mathbb{R}^{d\times d}$ are learnable weight matrices, and $\mathbf{b}^{(r1)}, \mathbf{b}^{(r2)}$ are bias vectors. $\mathbf{h}_i^{(r)}$ is the final output of one ASTGab for the AST node $i$.

\subsection{Predicting Next AST Nodes and Optimization}
\label{sec:pred_next}

As the right-most node has the most useful information of the next node, we use a soft-attention mechanism to incorporate the relevance of each node to the right-most node when generating the global representation of the AST graph: 
\begin{equation}
\small
\begin{aligned}
\beta_{i,n_j} &= \mathbf{z}^{(t1)}\sigma\Big(\mathbf{W}^{(t1)}_1\mathbf{h}_i^{(r)}+\mathbf{W}^{(t1)}_2\mathbf{h}_{n_j}^{(r)}+\mathbf{b}^{(t1)}\Big)\\
\mathbf{s}_j &= \sum_{i \in \mathcal{G}_j} \beta_{i,n_j}\mathbf{h}_i^{(r)}
\end{aligned}
\end{equation}
where $\mathcal{G}_j$ is the set of all unique AST nodes in the AST graph $j$, and $\mathbf{h}_{n_j}^{(r)}$ is the output from ASTGab(s) for the right-most node of the AST graph $j$. $\mathbf{W}_1^{(t1)}, \mathbf{W}_2^{(t1)}\in \mathbb{R}^{d\times d}$ are weight matrices, $\mathbf{b}^{(t1)}$ is a bias vector, $\sigma$ is the sigmoid function, $\mathbf{z}^{(t1)}\in \mathbb{R}^{d}$ is a learnable parameter vector, and $\mathbf{s}_j$ indicates the global representation of the AST graph $j$.

The global representation $\mathbf{s}_j$ is then projected to the value/type vocabulary space followed by softmax to generate the value/type probability distribution for the next node:
\begin{equation}
\small
\label{eq:pred}
\begin{aligned}
\mathbf{\bar{y}}_j^{(v)}&=\mathbf{W}^{(v)}\mathbf{s}_j+\mathbf{b}^{(v)},\,\,\mathbf{\hat{y}}_j^{(v)}=softmax\big(\mathbf{\bar{y}}_j^{(v)}\big)\\
\mathbf{\bar{y}}_j^{(t)}&=\mathbf{W}^{(t)}\mathbf{s}_j+\mathbf{b}^{(t)},\,\,\mathbf{\hat{y}}_j^{(t)}=softmax\big(\mathbf{\bar{y}}_j^{(t)}\big)
\end{aligned}
\end{equation}
where $\mathbf{W}^{(v)}\in \mathbb{R}^{V\times d},\mathbf{W}^{(t)}\in \mathbb{R}^{T\times d}$ are parameters, $\mathbf{b}^{(t)}\in\mathbb{R}^{V}$ and $\mathbf{b}^{(t)}\in\mathbb{R}^{T}$ are bias vectors, and $V$/$T$ is the vocabulary size of node value/type. $\mathbf{\hat{y}}_j^{(v)}\in\mathbb{R}^{V}$ and $\mathbf{\hat{y}}_j^{(t)}\in\mathbb{R}^{T}$ are the predicted value and type probability distributions of the next node to the partial AST corresponding to $j$.

Cross-entropy loss can be used for the optimization:
\begin{equation}
\mathcal{L}_v=-\sum_{j\in \mathcal{AG}}\mathbf{{y}}_j^{(v)}\log(\mathbf{{\hat{y}}}_j^{(v)}),\,\,\,\,\mathcal{L}_t=-\sum_{j\in \mathcal{AG}}\mathbf{{y}}_j^{(t)}\log(\mathbf{{\hat{y}}}_j^{(t)}),
\end{equation}
where $\mathcal{L}_v$ and $\mathcal{L}_t$ are the loss functions for value prediction and type prediction, respectively. $\mathcal{AG}$ is the set of all AST graphs generated from the data (i.e., one AST graph for one partial AST). $\mathbf{y}_j^{(v)}$ and $\mathbf{y}_j^{(t)}$ are two one-hot encodings of the ground-truth value and type for the next node, respectively. 

There are two prediction tasks in code completion, and two paradigms of training exist: (1) Train two models independently~\citep{LiWLK18}. One uses $\mathcal{L}_v$ for value prediction and the other uses $\mathcal{L}_t$ for type prediction. (2) Leverage multi-task learning~\citep{abs-2004-13379}, and train one model with a global loss function~\citep{LiuWSGS16,LiuLWXFJ20}: $\mathcal{L}=w_v\mathcal{L}_v+w_t\mathcal{L}_t$, where $w_v$ and $w_t$ are task weights. We choose the second training method. The reason is that value and type are two related attributes in code completion where the type can serve as a constraint to the value, and vice versa~\citep{LiuLWXFJ20}. Training two relevant tasks jointly via multi-task learning can benefit both tasks. Nevertheless, previous multi-task learning based code completion methods~\citep{LiuWSGS16,LiuLWXFJ20} treat two tasks equally (i.e., fix $w_v=w_t=1$) during optimization, while the two tasks may have different and changing converge speeds in multi-task learning~\citep{ChenBLR18}. Task imbalance will impede proper training because they manifest as imbalances between backpropagated gradients. However, the cost for manually tuning and updating task weights is unaffordable. To alleviate this issue, we use the idea of \emph{uncertainty}~\citep{KendallGC18} to automatically weigh two tasks during optimization without the need to tune task weights. As a result, the joint loss\footnote{Its derivation can be found in the appendix.} used by \ours is: 
\begin{equation}
\small
\label{eq:join_loss}
\begin{aligned}
\mathcal{L} &\approx \frac{1}{\theta^2}\mathcal{L}_{v} + \frac{1}{\tau^2}\mathcal{L}_{t} + \log\theta + \log\tau\\
&= \exp(-2\theta')\cdot\mathcal{L}_{v}+\exp(-2\tau')\cdot\mathcal{L}_{t} + \theta' + \tau',
\end{aligned}
\end{equation}
where $\theta$ and $\tau$ are learnable scalars. Their magnitudes indicate how ``uniform'' the discrete distributions are, which is related to the \emph{uncertainty} as measured in entropy~\citep{KendallGC18}. We let $\theta' = \log \theta, \tau' = \log \tau$, and train \ours to learn $\theta'$ and $\tau'$ instead of unconstrained $\theta$ and $\tau$ for the numerical stability. The reason is that $\frac{1}{\theta^{2}}$ and $\frac{1}{\tau^{2}}$ may encounter the overflow error for very small $\theta$ and $\tau$, and $\log \theta$ and $\log \tau$ will have the math domain error for nonpositive $\theta$ and $\tau$.
$\theta'$ and $\tau'$ are automatically learned parameters and can be interpreted as the task weights in multi-task learning~\citep{KendallGC18}.  
This way, we avoid manually setting the task weights and the two tasks are automatically balanced in multi-task learning so that \ours can provide accurate predictions for both tasks. All the parameters of \ours including $\theta'$ and $\tau'$ can be updated by gradient descent based methods.


\section{Experiments}

\begin{table*}[t]
\centering
\scalebox{0.78}{
\begin{tabular}{lcccccccccccc}
\hline
                  & \multicolumn{2}{c}{JS1k}                                                                                                                  & \multicolumn{2}{c}{JS10k}                                                                                                                 & \multicolumn{2}{c}{JS50k}                                                                                                                 & \multicolumn{2}{c}{PY1k}                                                                                                                   & \multicolumn{2}{c}{PY10k}                                                                                                                 & \multicolumn{2}{c}{PY50k}                                                                                                                 \\ \cline{2-13} 
                  & value                                                               & type                                                                & value                                                               & type                                                                & value                                                               & type                                                                & value                                                                & type                                                                & value                                                               & type                                                                & value                                                               & type                                                                \\ \hline
VanillaLSTM       & 53.19\%                                                             & 69.52\%                                                             & 58.04\%                                                             & 71.16\%                                                             & 59.70\%                                                             & 72.08\%                                                             & 49.99\%                                                              & 68.08\%                                                             & 52.67\%                                                             & 68.86\%                                                             & 53.66\%                                                             & 69.09\%                                                             \\
ParentLSTM        & 56.45\%                                                             & 71.99\%                                                             & 61.54\%                                                             & 73.46\%                                                             & 63.39\%                                                             & 74.24\%                                                             & 52.57\%                                                              & 70.10\%                                                             & 55.87\%                                                             & 76.25\%                                                             & 56.93\%                                                             & 71.00\%                                                             \\
PointerMixtureNet & 56.49\%                                                             & 71.95\%                                                             & 62.33\%                                                             & 74.28\%                                                             & 64.14\%                                                             & 76.01\%                                                             & 52.98\%                                                              & 69.98\%                                                             & 56.91\%                                                             & \textbf{76.94\%}                                                    & 57.22\%                                                             & 70.91\%                                                             \\
Transformer       & 58.40\%                                                             & \textbf{73.29\%}                                                    & \textbf{63.93\%}                                                    & \textbf{74.78\%}                                                    & 65.31\%                                                             & 75.89\%                                                             & 53.49\%                                                              & 70.63\%                                                             & 57.52\%                                                             & 71.45\%                                                             & 59.05\%                                                             & 71.91\%                                                             \\
Transformer-XL    & \textbf{59.23\%}                                                    & 72.11\%                                                             & 62.82\%                                                             & 74.09\%                                                             & \textbf{66.41\%}                                                    & \textbf{76.23\%}                                                    & \textbf{55.13\%}                                                     & \textbf{72.45\%}                                                    & \textbf{58.21\%}                                                    & 73.19\%                                                             & \textbf{60.00\%}                                                    & \textbf{72.42\%}                                                    \\ \hline
\ours             & \textbf{\begin{tabular}[c]{@{}c@{}}62.79\%\\ (6.01\%)\end{tabular}} & \textbf{\begin{tabular}[c]{@{}c@{}}75.72\%\\ (3.32\%)\end{tabular}} & \textbf{\begin{tabular}[c]{@{}c@{}}66.69\%\\ (4.32\%)\end{tabular}} & \textbf{\begin{tabular}[c]{@{}c@{}}78.55\%\\ (5.04\%)\end{tabular}} & \textbf{\begin{tabular}[c]{@{}c@{}}68.19\%\\ (2.68\%)\end{tabular}} & \textbf{\begin{tabular}[c]{@{}c@{}}80.14\%\\ (5.13\%)\end{tabular}} & \textbf{\begin{tabular}[c]{@{}c@{}}61.92\%\\ (12.32\%)\end{tabular}} & \textbf{\begin{tabular}[c]{@{}c@{}}76.71\%\\ (5.88\%)\end{tabular}} & \textbf{\begin{tabular}[c]{@{}c@{}}63.24\%\\ (8.64\%)\end{tabular}} & \textbf{\begin{tabular}[c]{@{}c@{}}80.90\%\\ (5.15\%)\end{tabular}} & \textbf{\begin{tabular}[c]{@{}c@{}}64.22\%\\ (7.03\%)\end{tabular}} & \textbf{\begin{tabular}[c]{@{}c@{}}75.31\%\\ (3.99\%)\end{tabular}} \\ \hline
\end{tabular}
}
\caption{Results on six datasets. Results of \ours and the best baselines are in bold. The percentages in brackets indicate the improvements of \ours over the best baselines.}
\label{tab:result}
\end{table*}

\subsection{Experimental Setup}

\subsubsection{Data.}
We choose two benchmark datasets\footnote{\url{https://www.sri.inf.ethz.ch/research/plml}} JavaScript (JS) and Python (PY) used in previous studies~\citep{LiWLK18,LiuLWXFJ20}. Each dataset contains 150,000 program files and their corresponding ASTs. We use the official train/test split, i.e., 100,000 in the training set and 50,000 in the test set.

We follow the method of \citet{LiWLK18} for data preprocessing.
We first flatten ASTs into sequences using pre-order
depth-first traversal. JS and PY have 95 and 330 types, respectively. We then divide each program into segments consisting of 50 consecutive AST nodes. Since the number of values is too large, we keep $K$ most frequent values in the training data to construct the value vocabulary where $K$=\{1,000, 10,000, 50,000\}. Values which are not included in the value vocabulary are replaced with \texttt{UNK}. For non-leaf nodes, \texttt{EMPTY} is used as the value. This way, we have 6 datasets: JS1k, JS10k, JS50k, PY1k, PY10k and PY50k. For each flattened sequence with $l$ AST nodes, we predict the value/type of the $r$-th ($2\le r \le l$) AST node based on its preceding $r-1$ nodes~\citep{LiWLK18,LiuLWXFJ20}.

\subsubsection{Baselines.}
We compare \ours with several state-of-the-art code completion methods. It is worth pointing out that some of them are simultaneously proposed in different papers with a slight difference in the design of the prediction layer. We group methods with similar designs as one baseline and describe the prediction layer we use. These baselines include:
\begin{itemize}[topsep=2pt,itemsep=0pt,leftmargin=10pt]
	\item \textbf{VanillaLSTM}~\citep{LiuWSGS16,LiWLK18,SvyatkovskiyZFS19} adopts LSTM to extract AST node features from flattened AST sequences. The inputs to each LSTM cell are previous hidden state and the concatenation of the type and value embeddings of the current AST node. The last output hidden state is fed to a prediction layer consisting of a single-layer feedforward neural network followed by softmax.
	\item \textbf{ParentLSTM}~\citep{LiWLK18} is an improved version of VanillaLSTM. It computes the attention weights between the current hidden state and all the previous hidden states within a context window to produce a context vector. The input to the prediction layer is the concatenation of the current hidden state, the context vector and the hidden state of the parent node of the current node.
	\item \textbf{PointerMixtureNet}~\citep{BhoopchandRBR16, LiWLK18} improves ParentLSTM using Pointer Network~\citep{VinyalsFJ15}. When predicting, it chooses the value/type for the next node from either a predefined global vocabulary or the context window according to the produced probability from the model.
	\item \textbf{Transformer}~\citep{abs-2003-13848,abs-2005-08025} improves VanillaLSTM by replacing LSTM with Transformer~\citep{VaswaniSPUJGKP17}.
	\item \textbf{Transformer-XL}~\citep{LiuLWXFJ20} adopts an improved Transformer architecture~\citep{DaiYYCLS19} to model the flattened AST sequence and the path from the predicting node to the root in the AST is fed to a BiLSTM to capture the structural information. The naive multi-task learning is used, and task weights are fixed to be equal.
\end{itemize} 
Transformer-XL and \ours (and its variants) adopt multi-task learning for training. For other methods, value prediction and type prediction are trained independently. 

\subsubsection{Hyper-parameters.}
For a fair comparison, we use 128 as the embedding size, hidden size and batch size for all methods. All methods are optimized with Adam~\citep{KingmaB14} using an initial learning rate of 0.001. For VanillaLSTM, ParentLSTM and PointerMixtureNet, the learning rate is multiplied by 0.6 after each epoch, the gradient norm is clipped to 5, and the size of context windows is set to 50 as suggested by~\citet{LiWLK18}. For Transformer based methods, we search the settings of heads and layer number in 1-6 and 1-8, respectively. Then, their best results are reported. Following \citet{LiWLK18}, for methods using parent-child relations, the hidden state of one parent node will be replaced with the hidden state of the immediate predecessor to the current node, if the parent node and the current node are not in the same segment. By default, we use 2 ASTGabs and 4 heads in \ours and its variants, but we also report the impacts of varying these hyper-parameters in Sec.\ref{sec:er}.

\subsubsection{Metrics.} We use accuracy as the evaluation metric for the code completion task~\citep{LiWLK18}. It shows the proportion of correctly predicted values/types for next AST nodes.

\subsection{Experimental Results}
\label{sec:er}

\begin{table}[t]
\centering
\scalebox{0.85}{
\begin{tabular}{lcccc}
\hline
\multicolumn{1}{l}{}      & \multicolumn{2}{c}{JS50k}           & \multicolumn{2}{c}{PY50k}           \\ \cline{2-5} 
\multicolumn{1}{l}{}      & value            & type             & value            & type             \\ \hline
\oursg                    & 66.20\%          & 75.97\%          & 62.05\%          & 73.22\%          \\
\oursn                    & 67.94\%          & 78.69\%          & 63.15\%          & 74.20\%          \\
\oursb                    & 65.27\%          & 75.10\%          & 59.22\%          & 71.45\%          \\
\oursp                    & 66.73\%          & 76.11\%          & 63.05\%          & 74.00\%          \\
\oursr                    & 66.76\%          & 76.31\%          & 62.35\%          & 73.64\%          \\ 
\oursng                   & 37.36\%          & 49.66\%          & 30.36\%          & 45.60\%          \\
\oursgs                   & 67.44\%          & 78.23\%          & 62.45\%          & 73.79\%          \\
\ourspe                   & 67.98\%          & 79.90\%          & 62.01\%          & 73.17\%          \\ \hline
\multicolumn{1}{l}{\ours} & \textbf{68.19\%} & \textbf{80.14\%} & \textbf{64.22\%} & \textbf{75.31\%} \\ \hline
\end{tabular}
}
\caption{Comparisons among variants of \ours.}
\label{tab:ablation}
\vspace{-10pt}
\end{table}

We run each method 3 times and report the average results. We conduct the Wilcoxon signed-rank test to test whether the improvements of \ours are statistically significant and all the p-values are less than 0.01.
We now analyze the results to answer several research questions:
 
\subsubsection{RQ1: How does \ours perform compared to baselines?} Tab.~\ref{tab:result} shows all methods' performance of both value prediction and type prediction on 6 datasets. We can see that \ours consistently surpasses the best baselines, by 2.68\%-12.32\% in value prediction and 3.32\%-5.88\% in type prediction. \ours also outperforms VanillaLSTM, which is used in Visual Studio Code IDE~\citep{SvyatkovskiyZFS19}, by 7.03\%-23.86\% in value prediction and 8.92\%-17.48\% in type prediction.  This demonstrates \ours's effectiveness.

\begin{figure}[!t]
\centering
\begin{subfigure}{.5\columnwidth}
  \centering
  \includegraphics[width=0.97\linewidth]{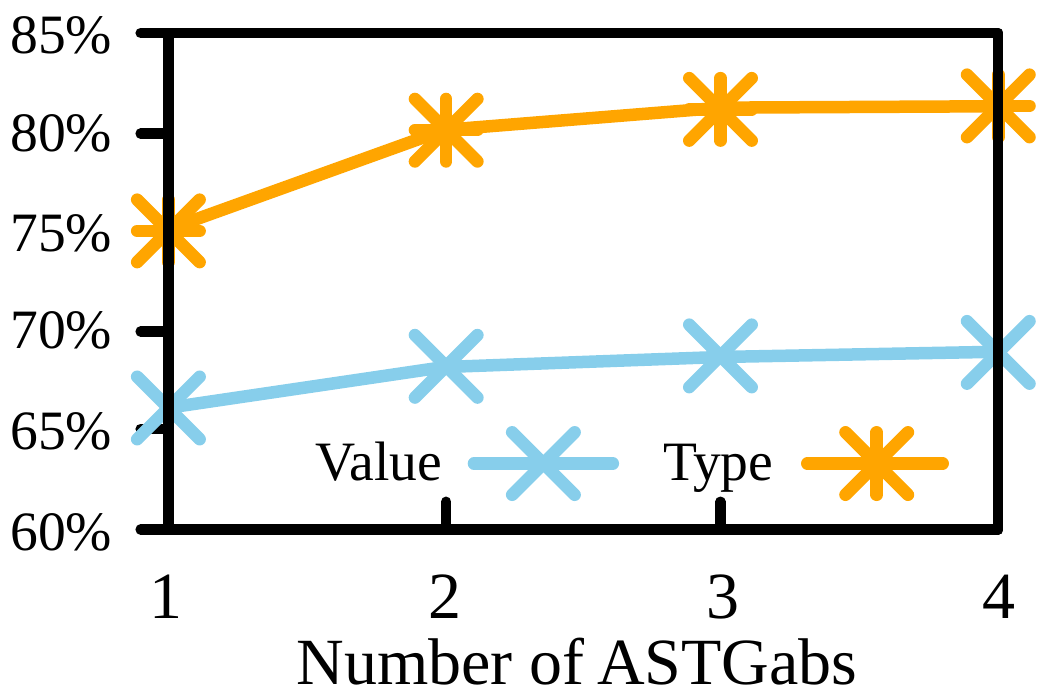}
\end{subfigure}%
\begin{subfigure}{.5\columnwidth}
  \centering
  \includegraphics[width=0.97\linewidth]{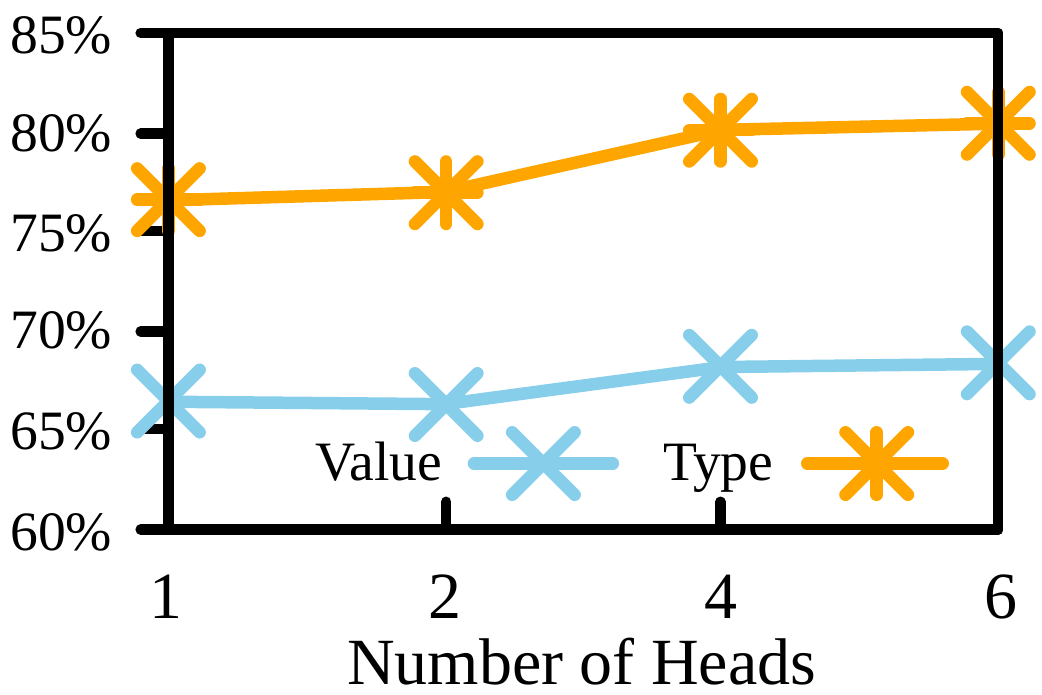}
\end{subfigure}
\caption{Impacts of hyper-parameters on \ours for JS50k. Left: Varying numbers of ASTGabs when 4 heads are used. Right: Varying numbers of heads when 2 ASTGabs are used.}
\label{fig:parameters}
\end{figure}

\begin{figure*}[t]
\begin{center}
\includegraphics[width=1\textwidth]{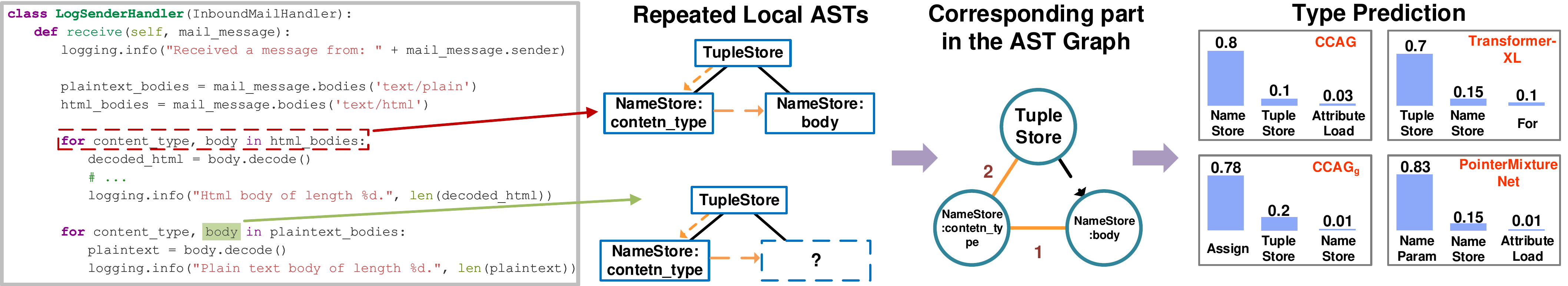}
\caption{A code completion example. The type of the next AST node \texttt{NameStore:body} is predicted by four methods. The orange lines illustrate the traversal for flattening the AST or the node-node edges. The black lines show parent-child edges.}
\label{fig:case_study}
\end{center}
\end{figure*}

\subsubsection{RQ2: Does each component of \ours contribute to the improvement of accuracy?} 
We conduct an ablation study to investigate whether each component of \ours contributes to its performance. Tab.~\ref{tab:ablation} reports the results of the following variants of \ours on JS50k and PY50k:
\begin{itemize}[topsep=2pt,itemsep=0pt,leftmargin=10pt]
\item \textbf{\oursg} models the original partial AST (instead of the flattened AST) as a graph, following the idea of \citet{AllamanisBK18,BrockschmidtAGP19}. The representation of each AST node is the concatenation of its type and value embeddings. 
\item \textbf{\oursn} fixes task weights in multi-task learning.
\item \textbf{\oursb} directly models the original partial AST as a graph, fixes task weights in multi-task learning and does not include PCAT and the residual connection.
\item \textbf{\oursp}, \textbf{\oursr}, \textbf{\oursng}, \textbf{\oursgs} and \textbf{\ourspe} do not include PCAT, the residual connection, NGAT, GSAT and the position embeddings, respectively.
\end{itemize}
Except the above differences, other parts of these variants are the same as \ours.
From Tab.~\ref{tab:ablation}, we can see that \oursg does not perform as well as \ours, which confirms the superiority of our design to construct AST graphs based on flattened ASTs instead of original partial ASTs. 
\oursn has worse accuracy than \ours, showing that using the uncertainty based method to automatically weigh the two tasks in multi-task learning can provide better results. 
\oursb only includes the core parts NGAT and GSAT, of which the idea comes from GNN (e.g., GAT~\citep{VelickovicCCRLB18}). \oursb has comparable performance to the best baselines in Tab.~\ref{tab:result}. Nevertheless, it performs worse than other variants with more components, showing that using more sophisticated neural networks like GNN is not the only reason for the superiority of \ours.
All of \oursp, \oursr, \oursng, \oursgs and \ourspe perform worse than \ours, which verifies that all components contribute to the effectiveness of \ours. \oursng shows the worst result among all variants. The reason is that ASTGab is based on GNN which updates node representations using information from the neighborhood. Hence, NGAT which aggregates information from neighbors is the most important component in ASTGab to implement GNN, and ASTGab without NGAT (i.e., \oursng) does not work at all.

\subsubsection{RQ3: What are the impacts when changing the hyper-parameters of \ours?} We conduct a study of the impacts of hyper-parameters on \ours. We show part of the results on JS50k in Fig.~\ref{fig:parameters}. For other settings and datasets, similar trends exist. Specifically, increasing the number of ASTGabs improves accuracy, but using more than 2 ASTGabs will not bring a significant gain. Besides, using more heads instead of a single head brings a noticeable improvement of accuracy. However, employing more than 4 heads does not improve accuracy too much.

\subsubsection{Case Study.} Fig.~\ref{fig:case_study} shows a program file in GitHub which is used to illustrate how the incoming email is handled in Google Cloud. It is also contained in the test set of PY. \texttt{NameStore:body} highlighted with green is the next node to predict. We provide the probabilities of the top-$3$ types predicted by four methods. \oursg gives high probabilities to the types of the first-order neighbors of the next node in the AST graph based on the original partial AST. Transformer-XL mainly gives high probabilities to the types of nodes along the path from the predicting node to the root in the original partial AST, and PointerMixtureNet gives high probabilities to the frequent types. Differently, \ours successfully captures the repetitive pattern (the repeated local ASTs) via the corresponding part in our AST graph which contains the repetitive pattern in edges and edge weights. Given the key part in our designed AST graph, it is easy for \ours to predict the next type \texttt{NameStore}.


\section{Related Work}

The pioneering work for code completion~\citep{BruchMM09} adopts Best Matching
Neighbor algorithm, which is later improved by \citet{ProkschLM15} with
Bayesian Networks. \citet{HindleBSGD12} find that most software can be
processed by NLP techniques, and they use n-gram language model for code
completion.

Recently, deep learning techniques such as RNNs~\citep{RaychevVY14},
GRUs~\citep{abs-2004-05249}, 
LSTMs~\citep{LiuWSGS16,SvyatkovskiyZFS19,abs-2003-08080}, Transformer and its
variants~\citep{abs-2003-13848,abs-2005-08025,LiuLWXFJ20}, and multi-task
learning~\citep{LiuWSGS16,LiuLWXFJ20} have been introduced to code completion
and significantly improved the accuracy of code completion engines. Besides,
some works consider modeling more context information.
\citet{BhoopchandRBR16,LiWLK18} adopt Pointer Network~\citep{VinyalsFJ15} and
keep the context representations of previous tokens to enhance code
completion. \citet{LiWLK18} consider the parent-child relation in AST to
capture the structure of AST. \citet{LiuLWXFJ20} model the path from the next
node to the root in the AST to improve code completion. \citet{abs-2004-13651}
explore different neural architectures (e.g., CNNs) for better context
encoding.

In addition to code completion, there are some works studying learning general
program representations for different downstream
tasks~\citep{AllamanisBDS18,LeCB20}. Some of them extract
AST-node-level~\citep{WangLT16} or AST-path-level
features~\citep{0002ZLY18,AlonBLY19,AlonZLY19}. Other works model ASTs as
trees~\citep{MouLZWJ16,abs-1910-12306,ZhangWZ0WL19} or
graphs~\citep{NguyenN15,abs-2004-02843,AllamanisBK18,BrockschmidtAGP19}  and
generate program representations.


\section{Conclusion}
This paper illustrates how to model the flattened AST as an AST graph to provide intelligent code completion. Our proposed \ours uses ASTGab to capture different dependencies in the AST graph and the sub-tasks of code completion are automatically balanced in optimization using uncertainty. Experimental results on benchmark data have demonstrated the effectiveness of \ours. In the future, we plan to explore the potential of modeling other representations of code (e.g., data flow graphs) to improve \ours further.

\section*{Acknowledgments}
Hui Li was partially supported by the National Natural Science Foundation of China (No. 62002303, 62036004), the Natural Science Foundation of Fujian Province of China (No. 2020J05001) and China Fundamental Research Funds for the Central Universities (No. 20720200089).


\section*{Appendix}
\subsection*{Derivation of Eq.~\ref{eq:join_loss}}

Existing code completion methods, which use multi-task learning, trains one model with a global loss function~\citep{LiuWSGS16,LiuLWXFJ20}: 
\begin{equation}
\small
\mathcal{L}=w_v\mathcal{L}_v+w_t\mathcal{L}_t,
\end{equation}
where $\mathcal{L}_v$ and $\mathcal{L}_t$ are defined in Eq.~11 of our submission. $w_v$ and $w_t$ are task weights for value prediction and type prediction, respectively. $w_v$ and $w_t$ are fixed and set to be equal~\citep{LiuLWXFJ20}. However, the two tasks may have different and changing converge speeds during multi-task learning~\citep{ChenBLR18,abs-2004-13379}. Task imbalances will impede proper training because they manifest as imbalances between backpropagated gradients. However, the cost for manually tuning and updating task weights is unaffordable. To alleviate this issue, we use the idea of \emph{uncertainty}~\citep{KendallGC18} and automatically weigh the two tasks during optimization without the need to tune task weights.

To be specific, we adapt the \textit{scaled} version of the value prediction for the next node of the partial AST $j$:
\begin{equation}
\small
\label{eq:softmax}
Pr\big(y^{(v)}_j\big|\mathbf{\bar{y}}_j^{(v)},\theta\big)=softmax\Big(\frac{1}{\theta^2}\mathbf{\bar{y}}_j^{(v)}\Big),
\end{equation}
where $\mathbf{\bar{y}}_j^{(v)}$ is the unnormalized probability distribution for the value of the next AST node as defined in Eq.~10 of Sec.~\ref{sec:pred_next}. $\theta$ is a positive scalar. Eq.~\ref{eq:softmax} can be interpreted as a Boltzmann distribution (also called Gibbs distribution)~\citep{KendallGC18}. The magnitude of $\theta$ indicates how ``uniform'' the discrete distribution is, which is related to its \emph{uncertainty} as measured in entropy~\citep{KendallGC18}. 

Taking the negative logarithm of the likelihood of next node value being value $c$, we have:
\begin{equation}
\small
\label{eq:log_soft}
\begin{aligned}
&\mathcal{L}_{v}(y^{(v)}_{j}=c)=-\log softmax\big(y_j^{(v)}=c, \mathbf{\bar{y}}_j^{(v)}\big)\\
&=-\mathbf{\bar{y}}_j^{(v)}(c) + \log\sum_{k\in \mathcal{V}, k\neq c} \exp\big(\mathbf{\bar{y}}_j^{(v)}(k)\big),
\end{aligned}
\end{equation}
where $\mathcal{V}$ is the vocabulary of AST node value, and $\mathbf{\bar{y}}_j^{(v)}(c)$ is the $c$-th dimension of $\mathbf{\bar{y}}_j^{(v)}$. Using Eq.~\ref{eq:log_soft}, the log likelihood for the output being node value $c$ in Eq.~\ref{eq:softmax} can be derived as:
\begin{equation}
\small
\begin{aligned}
&\log Pr\big(y^{(v)}_j=c\big|\mathbf{\bar{y}}_j^{(v)},\theta\big)\\
=&\frac{1}{\theta^2}\mathbf{\bar{y}}_j^{(v)}(c)-\log\sum_{k\in \mathcal{V}, k\neq c} \exp\big(\frac{1}{\theta^2}\mathbf{\bar{y}}_j^{(v)}(k)\big)\\
=&-\frac{1}{\theta^2}\mathcal{L}_{v}(y_t=c)+\log \bigg( \sum_{k\in \mathcal{V}, k\neq c} \exp \big(\mathbf{\bar{y}}_j^{(v)}(k) \big) \bigg)^{\frac{1}{\theta^2}}\\
&-\log \sum_{k\in \mathcal{V}, k\neq c}\exp \big(\frac{1}{\theta^2}\mathbf{\bar{y}}_j^{(v)}(k)\big).\\
\end{aligned}
\end{equation}

Applying the approximation proposed by ~\citet{KendallGC18}: 
\begin{equation}
\small
\label{eq:k_app}
\frac{1}{\theta}\sum_{k\in \mathcal{V}, k\neq c}\exp\Big(\frac{1}{\theta^2}\mathbf{\bar{y}}_j^{(v)}(k)\Big)\approx \bigg(\sum_{k\in \mathcal{V}, k\neq c}\Big(\exp\big(\mathbf{\bar{y}}_j^{(v)}(k)\big)\Big)\bigg)^{\frac{1}{\theta^2}},
\end{equation}	
we have:
\begin{equation}
\small
\label{eq:value_pro}
\footnotesize
\begin{aligned}
&\log Pr\big(y^{(j)}_j=c\big|\mathbf{\bar{y}}_j^{(v)},\theta\big)\\
=&-\frac{1}{\theta^2}\mathcal{L}_{v}(y_j=c)-\log\frac{\sum_{k\in \mathcal{V}, k\neq c}\exp \big(\frac{1}{\theta^2}\mathbf{\bar{y}}_j^{(v)}(k)\big)}{\bigg( \sum_{k\in \mathcal{V}, k\neq c} \exp \big(\mathbf{\bar{y}}_j^{(v)}(k) \big) \bigg)^{\frac{1}{\theta^2}}}\\
\approx & -\frac{1}{\theta^2}\mathcal{L}_{v}(y_j^{(v)}=c) -\log\theta.
\end{aligned}
\end{equation}

Similarly, for type prediction, we can get: 
\begin{equation}
\small
\label{eq:type_pro}
\log Pr\big(y^{(t)}_j=e\big|\mathbf{\bar{y}}_j^{(t)},\tau \big)\approx -\frac{1}{\tau^2}\mathcal{L}_{t}(y_j^{(t)}=e) -\log\tau,
\end{equation}
where $\tau$ is a positive scalar, $e$ is a type, and $\mathbf{\bar{y}}_j^{(t)}$ is the unnormalized probability distribution for the type of the next AST node as defined in Eq.~10 of our submission.

Given Eq.~\ref{eq:value_pro} and Eq.~\ref{eq:type_pro}, the joint loss with ground-truth value and type for the next node being $c$ and $e$ can be derived as:
\begin{equation}
\small
\label{eq:join_loss1}
\begin{aligned}
&\mathcal{L}(y_j^{(v)}=c,\, y_j^{(t)}=e,\, \theta,\, \tau) =\\
&= - \log \Big(Pr(y_j^{(v)}=c \big| \mathbf{\bar{y}}_j^{(v)},\theta)\cdot Pr(y_j^{(t)}=e \big| \mathbf{\bar{y}}_j^{(t)},\tau) \Big)\\
&\approx \frac{1}{\theta^2}\mathcal{L}_{v}(y_j^{(v)}=c) + \frac{1}{\tau^2}\mathcal{L}_{t}(y_j^{(t)}=e) + \log\theta + \log\tau.\\
&= \exp(-2\theta')\cdot\mathcal{L}_{v}(y_j^{(v)}=c)+\exp(-2\tau')\cdot\mathcal{L}_{t}(y_j^{(t)}=e) \\
&\,\,\,\,\,\,+ \theta' + \tau'.
\end{aligned}
\end{equation}

Then, taking all the next nodes into consideration, we have the join loss (i.e., Eq.~\ref{eq:join_loss} in Sec.~\ref{sec:pred_next}):
\begin{equation}
\small
\label{eq:join_loss2}
\begin{aligned}
\mathcal{L} &\approx \frac{1}{\theta^2}\mathcal{L}_{v} + \frac{1}{\tau^2}\mathcal{L}_{t} + \log\theta + \log\tau\\
&= \exp(-2\theta')\cdot\mathcal{L}_{v}+\exp(-2\tau')\cdot\mathcal{L}_{t} + \theta' + \tau'.
\end{aligned}
\end{equation}

In the last transition of Eqs.~\ref{eq:join_loss1} and~\ref{eq:join_loss2}, we let $\theta' = \log \theta$ and $\tau' = \log \tau$, and train the model to learn $\theta'$and $\tau'$ instead of the unconstrained scalars $\theta$ and $\tau$. This is for the numerical stability since $\frac{1}{\theta^{2}}$ and $\frac{1}{\tau^{2}}$ may encounter the overflow error for very small $\theta$ and $\tau$, and $\log \theta$ and $\log \tau$ will have the math domain error for nonpositive $\theta$ and $\tau$.

$\theta'$ and $\tau'$ are automatically learned parameters and can be interpreted as the task weights in multi-task learning~\citep{KendallGC18}. They are regularized by the last two terms in Eq.~\ref{eq:join_loss1} to prevent overfitting. This way, we avoid manually setting the task weights and the two tasks are automatically balanced in multi-task learning so that \ours can provide accurate predictions for both tasks. All the parameters of \ours including $\theta'$ and $\tau'$ can be updated by gradient descent based methods.

\clearpage

\bibliography{ref}

\end{document}